\documentclass[aps,prb,twocolumn,superscriptaddress,showpacs]{revtex4}
\usepackage{graphicx}
\usepackage{epsfig}
\usepackage{bm}
\usepackage{times}
\usepackage{amssymb}
\usepackage{color}

\begin{document}

\title{Raman scattering in a Heisenberg {\boldmath $S=1/2$} antiferromagnet on the triangular
lattice}

\author{Natalia  Perkins}
\affiliation{Department of Physics, University of Wisconsin-Madison,
Madison, WI 53706, USA}
\affiliation{Institute for Theoretical Physics,
Technical University Braunschweig, Mendelssohnstr. 3, 38106
Braunschweig, Germany}

\author{Wolfram Brenig}
\affiliation{Institute for Theoretical Physics,
Technical University Braunschweig, Mendelssohnstr. 3, 38106
Braunschweig, Germany}

\date{\today}

\begin{abstract}
We investigate two-magnon Raman scattering from the $S=1/2$
Heisenberg antiferromagnet on the triangular lattice, considering
both the effect of renormalization of the one-magnon spectrum  by
$1/S$ corrections and final-state magnon-magnon interactions. The
bare Raman intensity displays two peaks related to one-magnon
van-Hove singularities. We find that $1/S$ self-energy corrections
to the one-magnon spectrum strongly modify this intensity profile.
The central Raman-peak is significantly enhanced due to plateaus in
the magnon dispersion, the high frequency peak is suppressed due to
magnon damping, and the overall spectral support narrows
considerably. Additionally we investigate final-state interactions
by solving the Bethe-Salpeter equation to $O(1/S)$. In contrast to
collinear antiferromagnets, the non-collinear nature of the magnetic
ground state leads to an irreducible magnon scattering which is
retarded and non-separable already to lowest order. We show that
final-state interactions lead to a rather broad Raman-continuum
centered around approximately twice the 'roton'-energy. We also
discuss the dependence on the scattering geometry.
\end{abstract}

\maketitle

\section{Introduction}

Raman scattering is an  effective tool to study the excitation
spectrum of magnetic systems since the intensity of the
inelastically scattered light is directly related to the density of
singlet states at zero momentum. In local-moment magnets with well
defined magnon excitations this quantity is linked to the two-magnon
density of states. Therefore, magnetic Raman scattering plays an
important role in understanding the dynamics and interactions of
magnons in conventional spin systems
\cite{Elliot1963,Fleury1966,Parkinson1969,Fleury1970}. This is
particularly true  for the spin-$1/2$ square-lattice Heisenberg
antiferromagnet (HAF) of the high-T$_c$ superconductor parent
compounds,  where experimental
\cite{Lyons1988,Sugai1988,Sulewski1991} and theoretical
\cite{Singh1989,girvin,Chubukov1995,Nori1995} studies of the
magnetic correlations by Raman scattering  may provide important
insight into the energy scales relevant to the pairing mechanism
(for reviews see Refs. \cite{Blumberg1997,Devereaux2007}).

Raman scattering from HAFs can be understood in terms of the
Loudon-Fleury  (LF) processes  \cite{fleury}, in which two magnons
are simultaneously created by light absorption and emission. In the limit
of
large on-site Coulomb correlations $U$, the Hamiltonian  describing
these processes can be obtained   as a leading term of   the
expansion in $t/(U-\omega)$,  where $t$ is the nearest-neighbor (NN)
hopping, and $\omega$ is of the order of photon frequencies
\cite{Shastry1990}.

The Raman intensity of HAFs on hypercubic lattices with unfrustrated NN
exchange and collinear type of antiferromagnetic (AFM) order allows for a
straightforward semi-quantitative interpretation in terms of the LF
processes.  In fact, in real space, exchanging two NN spins of $S=1/2$
leads to an excitation with energy $\Omega \sim (z-1)J$, where $z$ is the
coordination number and $J$ is the AFM exchange energy. The reduction of
$\Omega/J$ by $-1$ is a consequence of the exchange link between the NN
sites and can be interpreted in terms of a two-magnon interactions in the
final state. In momentum space, the linear spin-wave theory yields
non-dispersive magnons along the the magnetic Brillouin zone (BZ) boundary,
leading to a square-root divergence of the bare two-magnon density of
states at $\Omega=zJ$. Inclusion of the final state magnon-magnon
interactions broadens the singularity and shifts it down to $\Omega \sim
2.9J$ \cite{Singh1989,girvin,Chubukov1995,Nori1995} in two dimensions,
which is consistent, both with the real-space result $\Omega=J(4-1)=3J$,
and with the experimentally observed Raman profile.

In contrast to conventional collinear HAFs, very little is known
theoretically about Raman scattering from frustrated HAFs. This is
intriguing, since the singlet spectrum is believed to be an
essential fingerprint of such magnets. The spin $S=1/2$ HAF on the
triangular lattice (THAF) with NN exchange interactions is a
prominent example of strongly frustrated spin systems. It has a
ground state with non-collinear $120^\circ$ degree ordering of the
spins. Due to this non-collinearity of the classical ground state,
nontrivial corrections to the spin-wave spectrum
 appear already to first order in $1/S$. It has been shown in
Refs.\cite{chubukov,chubukov06,mike} that $1/S$ corrections strongly modify
the form of the magnon dispersion of the triangular HAF. The resulting
dispersion  turns out to be almost flat in a wide range of momenta in which
it possesses shallow local minima, "rotons",  at the midpoint of the faces
of the hexagonal BZ. This differs strongly from the classical spin-wave
spectrum, which lacks such minima and flat zones. Similar results have been
 obtained in   series expansion  studies \cite{Zheng2006}.

Motivated by these recent findings, in this paper, we analyze the
Raman scattering from the THAF by $1/S$ expansion. This is
complementary to the recent analysis of Raman scattering on finite,
16 sites THAFs by means of exact diagonalization \cite{Vernay2007}.
First, our results show that the Raman intensity is very sensitive
to both $1/S$ corrections of the magnon spectrum and  the
magnon-magnon interactions in the final state. Moreover, we find
that the Loudon-Fleury process on the THAF leads to a Raman profile,
which is independent at $O(1/S)$ of the scattering geometry.

The manuscript is organized as follows. In section \ref{sec2}, we
review results from existing calculations \cite{chubukov,chubukov06}
of the one magnon excitations in the THAF to first order in $1/S$
needed for our study of the Raman spectra. In section \ref{sec3} we
consider the LFP to leading order in $1/S$. In section \ref{sec4} we
calculate the Raman spectrum on various levels of approximation in
$1/S$, i.e. bare, one-magnon renormalized, and  including final
state interactions, and show that Raman profile is  very sensitive
to the magnon-magnon interactions. We discuss our results in section
\ref{sec5}.

\section{Model}\label{sec2}

\begin{figure}
\centerline{\epsfig{file=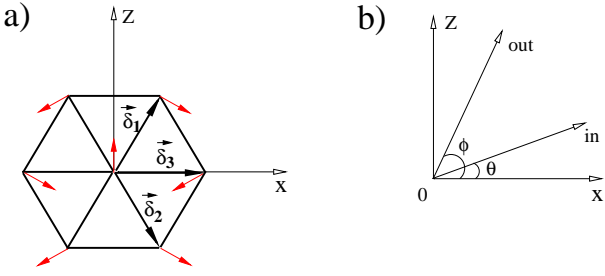,width=8cm}}
 \caption{a) Classical $120^{\circ}$ degree non-collinear spin
order on the triangular lattice.
 Basic vectors of triangular lattice:
${\vec \delta}_1=(\frac{1}{2},\frac{\sqrt{3}}{2})$, ${\vec
\delta}_2= (\frac{1}{2},-\frac{\sqrt{3}}{2})$ and  ${\vec
\delta}_3=(1,0)$. b) Definition of scattering angles
for LF vertex.} \label{geometry}
\end{figure}

The Hamiltonian of the THAF reads
\begin{equation}\label{eq1}
H=J\sum_{\langle ij\rangle}{\bf S}_i{\bf S}_j ,
\end{equation}
where ${\bf S}_i$ are spin$-1/2$ operators, $i$ refers to sites on
the triangular lattice, $\langle \rangle$ denotes NN summation, and
$J$ is the exchange interaction. The classical ground state of the
THAF \cite{Jolicoeur1989} is a non-collinear $120^\circ$ degree ordering
of spins which is shown in Fig.\ref{geometry}a).  To avoid the
complexity of a three-sublattice notation it is convenient to work
within a locally rotated frame of reference in which the magnetic
order is ferromagnetic. To achieve this we assume a gauge in which
the $(x,z)$-coordinates label the lattice plane and a uniform twist
with a pitch vector ${\bf Q}=(4\pi/3,0)$ is applied to the $y$-axis.
The laboratory frame-of-reference spin ${\bf S}_i$ is related to the
spin $\tilde{{\bf S}}_i$ in the rotated frame through
\begin{equation}\label{eq2}
{\bf S}_i =
\left[\begin{array}{ccc}
\sin(q_i) & -\cos(q_i) & 0 \\
0 & 0 & -1 \\
\cos(q_i) & \sin(q_i) & 0
\end{array}\right]^{-1}
\tilde{{\bf S}}_i,
\end{equation}
where $q_i=2\pi (2 l_i + m_i)/3$ and $(l_i,m_i)$ are integers
labeling the points on the triangular lattice, which is depicted in
Fig. \ref{geometry}a). In contrast to ${\bf S}_i$, the spin
$\tilde{{\bf S}}_i$ is amenable to a representation in terms of a
{\em single} Holstein-Primakoff boson field on all sites
\begin{eqnarray}\label{eq3}
\tilde{S}^z_i &=& S - a^+_ia^{\phantom{+}}_i\\\nonumber
\tilde{S}^+_i &=& (2S - a^+_ia^{\phantom{+}}_i)^{1/2}
a^{\phantom{+}}_i\\\nonumber \tilde{S}^-_i &=& a^{+}_i (2S -
a^+_ia^{\phantom{+}}_i)^{1/2}~.
\end{eqnarray}

Because we intend to study magnon interactions to first order in
 $1/S$, we need to expand the Hamiltonian in Eqn. (\ref{eq1}) up to
quartic order in the boson fields. We have
\begin{eqnarray}\label{eqn4}
H-E_0=3 J S (H_2+H_3+H_4)~,
\end{eqnarray}
\noindent
where $E_0=3 J S^2/2$ is the classical ground state energy and
\begin{eqnarray}
\label{h2}
H_2=\sum_{\bf k} A_{\bf k}
a_{\bf k}^{\dagger}a^{\phantom{\dagger}}_{\bf k}+
\frac{B_{\bf k}}{2}(a_{\bf k}^{\dagger}a_{-{\bf k}}^{\dagger}
+a^{\phantom{\dagger}}_{\bf k}a^{\phantom{\dagger}}_{-{\bf k}})
\phantom{a a a a a a a a a} && \\[.2cm]
\label{h3}
H_3= -i \sqrt{\frac{3}{8S}}
\sum_{{\bf k}_1,{\bf k}_2,{\bf k}_3}  (a_{{\bf k}_1}^{\dagger}
a_{{\bf k}_2}^{\dagger}a^{\phantom{\dagger}}_{{\bf k}_3}
-a_{{\bf k}_3}^{\dagger}a^{\phantom{\dagger}}_{{\bf
k}_2}a^{\phantom{\dagger}}_{{\bf k}_1})\times
\phantom{a a} && \\ \nonumber
(\bar{\nu}_{{\bf k}_1}+\bar{\nu}_{{\bf k}_2})
\delta_{{\bf k}_3,{\bf k}_1+{\bf k}_2} && \\[.2cm]
\label{h4}
H_4 =-\frac{1}{16S} \sum_{{\bf k}_1,{\bf k}_2,{\bf k}_3,{\bf k}_4}
\delta_{{\bf k}_3+{\bf k}_4,{\bf k}_1+{\bf k}_2} \,
a_{{\bf k}_1}^{\dagger}a_{{\bf k}_2}^{\dagger}
a^{\phantom{\dagger}}_{{\bf k}_3}a^{\phantom{\dagger}}_{{\bf k}_4}
\times \phantom{a} && \\
(4(\nu_{{\bf k}_1-{\bf k}_3}+\nu_{{\bf k}_2-{\bf k}_3})+
\nu_{{\bf k}_1}+ \nu_{{\bf k}_2}+\nu_{{\bf k}_3}+ \nu_{{\bf k}_4})) -
&& \nonumber \\
2\,\delta_{{\bf k}_1+{\bf k}_2+{\bf k}_3,{\bf k}_4}\,
(a_{{\bf k}_1}^{\dagger}
a_{{\bf k}_2}^{\dagger}
a_{{\bf k}_3}^{\dagger}
a^{\phantom{\dagger}}_{{\bf k}_4}
+a_{{\bf k}_4}^{\dagger}
a^{\phantom{\dagger}}_{{\bf k}_3}
a^{\phantom{\dagger}}_{{\bf k}_2}
a^{\phantom{\dagger}}_{{\bf k}_1}) && \nonumber \\
(\nu_{{\bf k}_1}+\nu_{{\bf k}_2}+\nu_{{\bf k}_3})~, &&
\nonumber
\end{eqnarray}
where momentum ${\bf k}$ is defined in the first  magnetic
BZ. We use the following notations:
\begin{eqnarray}
 A_{\bf k}=1+\nu_{\bf k}/2,~~ B_{\bf k}=-3\nu_{\bf k}/2~,
\end{eqnarray}
\noindent and the momentum dependent functions are
\begin{eqnarray}
\nu_{\bf k}&=&\frac{1}{3}(\cos k_x
+2\cos\frac{k_x}{2}\cos\frac{k_y\sqrt{3}}{2})~,\\
\bar{\nu}_{\bf k}&=& \frac{2}{3}\sin\frac{k_x}{2}(\cos k_x
-\cos\frac{k_y\sqrt{3}}{2})~.
\label{nunu}
\end{eqnarray}
The expressions for $H_3$ and $H_4$ have been obtain first in Ref.
\cite{chubukov}. The essential difference between Eqns. (\ref{eqn4})
- (\ref{h4}) and a corresponding expansion around a Ne\'el state on
a hypercubic lattice is the occurrence of the term $H_3$, which is
present due to the non-collinearity of the classical ground state
configuration of the THAF. In the remainder of this paper we set the
scale of energy to $3J/2=1$, i.e. for $S=1/2$ the prefactor in Eqn.
(\ref{eqn4}) is unity.

\begin{figure}
\centerline{\epsfig{file=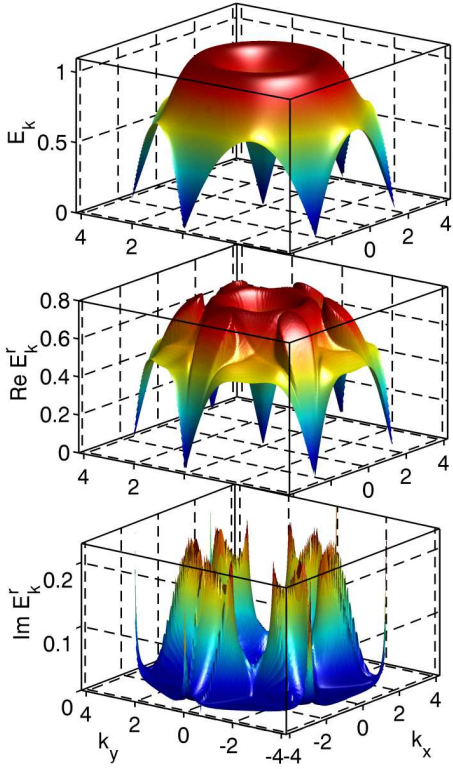,width=7cm}} \caption{One magnon
dispersion. Top: linear spin-wave dispersion $E_{\bf k}$ of from
Eqn. (\ref{ek}). Middle and bottom: real and imaginary part $Re(Im)
E^r_{\bf k}$ of one magnon dispersion to $O(1/S)$ from a solution of
Eqns. (10) - (12) of Ref. \cite{chubukov06} on a lattice of
$252\times 252$ ${\bf k}$-points with artificial line broadening of
$\eta =0.05 $.} \label{fig2}
\end{figure}

To proceed, we diagonalize the quadratic part of the Hamiltonian
$H_2$ by a Bogoliubov transformation to a set of magnon
quasiparticles
\begin{eqnarray}
\label{bt}
a^{\phantom{\dagger}}_{\bf k}&=&
u_{\bf k}c^{\phantom{\dagger}}_{\bf k}+
v_{\bf k}c_{-{\bf k}}^{\dagger}
\\\nonumber
a_{\bf k}^{\dagger}&=&
u_{\bf k}c_{\bf k}^{\dagger}+
v_{\bf k}c^{\phantom{\dagger}}_{-{\bf k}}~,
\end{eqnarray}
where $c^{(\dagger)}_{\bf k}$ are bosons, and the coherence
coefficients
\begin{eqnarray}
\label{uv} u_{\bf k}&=&
\sqrt{\frac{A_{\bf k}+E_{\bf k}}{2E_{\bf k}}}\\\nonumber
v_{\bf k}&=&
-\frac{B_{\bf k}}{|B_{\bf k}|}
\sqrt{\frac{A_{\bf k}- E_{\bf k}}{2E_{\bf k}}}~.
\end{eqnarray}
satisfy $u_{\bf k}^2-v_{\bf k}^2=1$. The Hamitonian $H_2$ in terms of
the Bogoliubov quasiparticles reads
\begin{eqnarray}
\label{h2diag} H_2=\sum_{\bf k} E_{\bf k}c_{\bf k}^{\dagger}c_{\bf
k}~,
\end{eqnarray}
and  the dispersion is given by
\begin{eqnarray}
\label{ek} E_{\bf k}=\sqrt{A_{\bf k}^2-B_{\bf
k}^2}=\sqrt{(1-\nu_{\bf k}) (1+2\nu_{\bf k})}~.
\end{eqnarray}
The magnon dispersion $E_{\bf k}$ is depicted in Fig. \ref{fig2}. It
vanishes at the center of the zone, $k_x=0, k_y=0$, where $\nu_{\bf
k}=1$ and at the corners of the BZ, where $\nu_{\bf k}=-1/2$. There
are two van-Hove singularities, i.e. at $E=3/(2\sqrt{2})\simeq 1.061$ from
the maximum energy and at $E=2/3\simeq 0.6667J$ from the zone-boundary.

To treat the interaction between magnons we need to express the
triplic and quartic part of the Hamiltonian, $H_3$ and $H_4$, in
terms of the quasiparticles $c^{(\dagger)}_{\bf k}$ using the
transformation of Eqn. (\ref{bt}). For the triplic part we obtain
\begin{eqnarray}\label{h3quas}
H_3 = \sum_{{\bf k},{\bf p}}\, \left(c_{{\bf k}}c_{{\bf
p}}^{\dagger}c_{{\bf k}-{\bf p}}^{\dagger} \, f\left({\bf k},{\bf
p}\right) + \right. \phantom{a a a a a a} && \nonumber \\\left.
c_{{\bf k}}c_{{\bf p}}^{\dagger}
c^{\phantom{\dagger}}_{-{\bf k}+{\bf p}}\,g\left({\bf k},{\bf p}\right)
+ O(c^\dagger_{\phantom{\bf k}}c^\dagger_{\phantom{\bf k}}
c^\dagger_{\phantom{\bf k}}+h.c.)
\right)~. &&
\end{eqnarray}
Terms with three creation(destruction) operators are present in
principle, but are not expressed explicitly for notational
simplicity. As will be come clear in section \ref{sec4} they play no
role in evaluating the magnon interactions within the Raman
response. Moreover
\begin{eqnarray}
f({\bf k},{\bf p}) &=&
i\sqrt{3}(\bar{\nu}_{{\bf p}}(u_{{\bf k}}u_{{\bf
{\bf k}}-{\bf p}}+v_{{\bf k}}v_{{\bf k}-{\bf p}})(u_{{\bf p}}+v_{{\bf p}})-
\nonumber \\
&& \bar{\nu}_{{\bf k}}(u_{{\bf k}}+v_{{\bf k}})(u_{{\bf p}}v_{{\bf
{\bf k}}-{\bf p}}+u_{{\bf k}-{\bf p}}v_{{\bf p}})+
\nonumber \\
&& \bar{\nu}_{{\bf k}-{\bf p}}(u_{{\bf k}-{\bf p}}+v_{{\bf k}-{\bf p}}
)(u_{{\bf k}}u_{{\bf p}}+v_{{\bf k}}v_{{\bf p}}))~,
\label{fkp}\\[.2cm]
g({\bf k},{\bf p})&=&
i\sqrt{3}(\bar{\nu}_{{\bf p}}(u_{{\bf
{\bf k}}-{\bf p}}v_{{\bf k}}+u_{{\bf k}}v_{{\bf k}-{\bf p}})(u_{{\bf
{\bf p}}}+v_{{\bf p}})+
\nonumber \\
&&\bar{\nu}_{{\bf k}-{\bf p}}(u_{{\bf k}-{\bf p}}+v_{{\bf
{\bf k}}-{\bf p}})(u_{{\bf k}}u_{{\bf p}}+v_{{\bf k}}v_{{\bf p}})-
\nonumber \\
&&\bar{\nu}_{{\bf {\bf k}}}(u_{{\bf k}}+v_{{\bf {\bf k}}})(u_{{\bf
k}-{\bf p}}u_{{\bf p}}+v_{{\bf k}-{\bf p}}v_{{\bf p}}))~.
\label{gkp}
\end{eqnarray}
For the quartic part we obtain
\begin{eqnarray}\label{h4quas}
H_4=-\frac{1}{16S}
\sum_{{\bf k},{\bf p}} \,
h({\bf k},{\bf p}) \,
c_{{\bf k}}^{\phantom{\dagger}}
c_{-{\bf k}}^{\phantom{\dagger}}
c_{{\bf p}}^{\dagger}
c_{-{\bf p}}^{\dagger} &&\nonumber \\
+ O( c^\dagger_{\phantom{\bf k}} c^\dagger_{\phantom{\bf k}}
c^\dagger_{\phantom{\bf k}} c^\dagger_{\phantom{\bf k}} +
c^\dagger_{\phantom{\bf k}} c^\dagger_{\phantom{\bf k}}
c^\dagger_{\phantom{\bf k}} c^{\phantom{\dagger}}_{\phantom{\bf k}}
+h.c.) &&~,
\end{eqnarray}
where again terms irrelevant for the Raman scattering are not
displayed explicitly and
\begin{eqnarray}\label{h4a}
h({\bf k},{\bf p}) =
2 ((u_{{\bf k}}^{2}u_{{\bf p}}^{2}+v_{{\bf k}}^{2}v_{{\bf p}}^{2})
(\nu_{{\bf k}}+4\nu_{{\bf k}-{\bf p}}+\nu_{{\bf p}})-3(u_{{\bf k}}^{2}+
&& \nonumber \\
v_{{\bf k}}^{2})u_{{\bf p}}v_{{\bf p}}(2\nu_{{\bf
k}}+\nu_{{\bf p}})
-3(u_{{\bf p}}^{2}+v_{{\bf p}}^{2})u_{{\bf k}}v_{{\bf k}}(\nu_{{\bf
k}}+2\nu_{{\bf p}})
\phantom{a a}&&  \nonumber \\
+4u_{{\bf k}}v_{{\bf k}}u_{{\bf p}}v_{{\bf p}}(2+\nu_{{\bf
k}}+\nu_{{\bf p}}+2\nu_{{\bf k}+{\bf p}})) \phantom{a a a a}&&~.
\end{eqnarray}

Eqns. (\ref{h3quas}) -- (\ref{h4a}) allow to construct all vertices
relevant to the final state two-magnon interactions in the Raman
scattering. Apart from that Eqns. (\ref{eqn4}) -- (\ref{nunu}) can
be used to  derive the one-magnon selfenergy to $O(1/S)$. This has
been done in Ref. \onlinecite{chubukov}, to which we refer the
reader for details. For the purpose of the present work it is
sufficient to employ Eqns. (10) - (12) from
Ref.\onlinecite{chubukov} to calculate the renormalized magnon
dispersion $E^r_{\bf k}$ to $O(1/S)$. Fig. \ref{fig2} (middle
and bottom panel) shows the result of such calculations. It is
evident that in the real part of the magnon energy, the interactions
lead to extended and almost flat regions with a shallow 'roton'-like
minimum along the BZ faces. Moreover, as the $Im E^r_{\bf k}$ almost
vanishes at these regions, the life time of a quasi-particle with
corresponding momenta is very large. On the other hand
quasi-particles with near-maximum energies are located in the
momentum regions of rather large damping.

\section{Loudon-Fleury Vertex}\label{sec3}

We use the framework of the Loudon-Fleury (LF) model for the
interaction of light with spin degrees of freedom for the
calculation of the two-magnon Raman scattering.
The LF vertex  has
the form
\begin{eqnarray}
 \begin{array}{l}
R=\sum_{i\delta}(\hat{\varepsilon}_{\rm in}\cdot {\bf\delta})
(\hat{\varepsilon}_{\rm out}\cdot {\bf\delta}) \tilde{{\bf
S}}_i\tilde{{\bf S}}_{i+\delta}~,
\end{array}
\label{loudonf}
\end{eqnarray}
where the polarizations $\hat{\varepsilon}_{\rm in}=\cos\theta \hat
x+\sin\theta \hat y$  and $\hat{\varepsilon}_{\rm out}=\cos\phi \hat
x+\sin\phi \hat y$ of the incoming  and the outgoing light are
determined by angles $\theta$ and $\phi$, defined with respect to
the $x$-axis. To derive the final form of the scattering LF vertex,
we first write spin operators in terms of  Holstein-Primakoff
bosonic $a$-operators (\ref{eq3}), and then  express the  latter in
terms of the boson quasi-particle operators $c$. We get the
following expression
\begin{eqnarray}
 \begin{array}{l}
R=\sum_k M_{\bf k}(c_{\bf k}c_{-{\bf k}}+ c_{\bf
k}^{\dagger}c_{-{\bf k}}^{\dagger}) \equiv r^- + r^+ ~,
\end{array}
\label{lf2}
\end{eqnarray}
where $M_{\bf k}$ is given by
\begin{eqnarray}
\begin{array}{ll}
M_{\bf k}=&(F_1(\theta,\phi)+F_2({\bf k},\theta,\phi))u_{\bf
k}v_{\bf k}-\\& \frac{3}{4}F_2({\bf k},\theta,\phi)(u_{\bf
k}^2+v_{\bf k}^2)~,
\end{array}
\end{eqnarray}
and we have introduced the following notations:
 \begin{eqnarray}
 \begin{array}{ll}
&F_1(\theta,\phi)=2S\sum_{\mu=1}^3f_{\mu}(\theta,\phi)~, \\[.2cm]
&F_2({\bf k},\theta,\phi)=2S(f_3(\theta,\phi)\cos k_x+\\&
f_1(\theta,\phi)\cos(\frac{k_x}{2}+\frac{\sqrt{3}}{2}k_y)
+f_2(\theta,\phi)\cos(\frac{k_x}{2}-\frac{\sqrt{3}}{2}k_y))~,\\[.2cm]
&f_{\mu}(\theta,\phi)=\hat{\varepsilon}_{\rm in}\cdot
\vec{\delta_{\mu}}~.
\end{array}
\label{fk1}
\end{eqnarray}
In principle the Raman vertex contains also $c^{\dagger}_{\bf
k}c_{\bf k}$  terms. However, at zero momentum and to lowest order
in $1/S$ these terms do not contribute to the  Raman response at
finite frequency, and we dropped them. Note that $R$ is explicitly
Hermitian.

\begin{figure}
\centerline{\epsfig{file=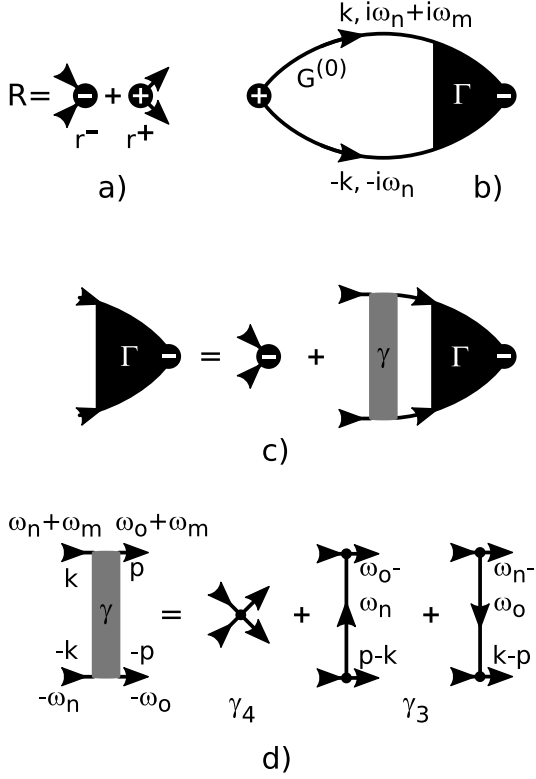,width=7cm}}
 \caption{Diagrams for the Raman intensity: a) Bare Raman vertex $R$ from
Eqn. (\ref{lf2}); b) Raman susceptibility (both  bare,  $G^0$, and
dressed, $G$, magnon propagators will be considered (see text). c)
The integral equation for the dressed Raman vertex $\Gamma$ in terms
of the irreducible magnon particle-particle (IPP) vertex $\gamma$.
d) Leading order $1/S$ contributions to the IPP vertex.}
\label{gamam}
\end{figure}

\section{Raman Intensity}\label{sec4}

We now calculate the Raman intensity including one- and
two-magnon renormalizations up to $O(1/S)$. The Raman intensity $I(\Omega)$
is obtained via Fermi's golden-rule from
\begin{equation}
I(\omega_n) = const\times Im [
\int_0^\beta d\tau\, e^{i \omega_m \tau}
\left\langle T_\tau (
R(\tau) R
) \right\rangle ]
\label{Fermi}
\end{equation}
by analytic continuation of the bosonic Matsubara frequencies
$\omega_m = 2\pi m T $ onto the real axis as $i\omega_m\rightarrow
\Omega+i\eta$, where $\Omega=\omega_{\rm in}-\omega_{\rm out}$
refers to the inelastic energy transfer by the photon, and  for the
remainder of this paper we assume the temperature $T=1/\beta$ to be
zero. The prefactor  '$const$' refers to some arbitrary units by
which the observed intensities are scaled.

The role of interactions is summarized in Fig. \ref{gamam}. Two
effects have to be distinguished, namely renormalizations of the
one-magnon propagators, i.e. $G^0\rightarrow G$, and vertex
corrections to the Raman intensity (final state interactions), i.e
$R\rightarrow \Gamma$.

All propagators are expressed in terms $c_{k}^{(\dagger)}$-type of
Bogoliubov particles. To orders higher than $O((1/S)^0)$ the
propagators of these particles are not diagonal i.e.  both normal
$G_{cc}({\bf k},\tau)=-\langle T_\tau ( c_{{\bf k}}(\tau) c_{{\bf
k}}^{\dagger})\rangle$  and anomalous $D_{cc}({\bf k},\tau)=-\langle
T_\tau ( c_{{\bf k}}(\tau) c_{-{\bf k}})\rangle$ propagators do
occur. However, anomalous propagators are smaller by one factor of
$1/S$ as compared to the normal propagators and, therefore,  can be
neglected. To first order in $1/S$, the normal propagators  read
$G({\bf k}, i\omega_n)=1/(i\omega_n-E^r_{\bf k})$, i.e. the
quasi-particle residue remains unity, and $E^r_{\bf k}$ is taken
from Eqns. (10) - (12) of Ref. \onlinecite{chubukov06}.

In order to evaluate Raman intensity (Eqn. (\ref{Fermi})) we have
to calculate the Raman susceptibility $\left\langle T_\tau(
R(\tau) R) \right\rangle$. In principle, the latter can contain
terms of type $\left\langle T_\tau( r^-(\tau) r^-)\right\rangle$,
with $r^\pm$ specified in Eqn. (\ref{lf2}) and Fig. \ref{gamam}~a).
However, these terms need at least one $O(1/S)$ interaction-event to
occur or require anomalous propagators, i.e., they are smaller by
one order of $1/S$ and will be dropped. In the following we consider
only $\left\langle T_\tau(r^{+}(\tau) r^-)\right\rangle
+\left\langle T_\tau (r^-(\tau) r^{+})\right\rangle$. By Hermitian
conjugation, it is sufficient to calculate $J(\tau)=\left\langle
T_\tau (r^-(\tau) r^+)\right\rangle$, which is depicted in Fig.
\ref{gamam}~b).

Fig. \ref{gamam}~b)  shows the two-particle reducible Raman vertex
$\Gamma({\bf k},\omega_n,\omega_m)$, which includes a series of
magnon-magnon interaction events. It satisfies the Bethe-Salpeter
equation expressed in terms of the two-particle irreducible vertex
$\gamma$, depicted in Fig. \ref{gamam}~c)
\begin{eqnarray}\label{bseqn}
\Gamma ({\bf k},\omega_n,\omega_m) =
r^-({\bf k}) +
\sum_{{\bf p},\omega_o}
\gamma ({\bf k},{\bf p},\omega_n,\omega_o)
&&\nonumber\\
G({\bf p},\omega_o+\omega_m)
G(-{\bf p},-\omega_o)
\Gamma ({\bf p},\omega_o,\omega_m)&&
\end{eqnarray}

In this work we consider only the leading order contributions in
$1/S$ to $\gamma$. They are shown in  Fig. \ref{gamam}~d). The
quartic vertex $\gamma_4({\bf k},{\bf p})$ is identical to the
two-particle-two-hole contribution from $H_4$ of
Eqns.(\ref{h4quas})-(\ref{h4a}) and reads
\begin{eqnarray}
 \begin{array}{l}
 \gamma_4({\bf k},{\bf
p})=-\frac{1}{2S}h({\bf k},{\bf p})
\end{array}
\label{gamma4f}
\end{eqnarray}
The two addends forming the irreducible vertex $\gamma_3 ({\bf
k},{\bf p},\omega_n,\omega_o)$ are assembled from $H_3$ and one
intermediate propagator, and can be written as
\begin{eqnarray}\label{gamma3a}
\gamma_3 ({\bf k},{\bf p},\omega_n,\omega_o)
=\frac{1}{2S}\sum_{{\bf k},{\bf p}}\,(
f({\bf k},{\bf p})g(-{\bf k},-{\bf p})\times
&&\nonumber\\
G^0({\bf k}-{\bf p},i\omega_o-i\omega_n)
c_{\bf k}^{\phantom{\dagger}}
c_{-{\bf k}}^{\phantom{\dagger}}
c_{\bf p}^{\dagger}
c_{-{\bf p}}^{\dagger}+
\phantom{a a a }&&\\
g({\bf k},{\bf p})f(-{\bf k},-{\bf p}) G^0({\bf p}-{\bf
k},i\omega_n-i\omega_o)) c_{\bf k}^{\phantom{\dagger}} c_{-{\bf
k}}^{\phantom{\dagger}} c_{\bf p}^{\dagger} c_{-{\bf
p}}^{\dagger}&&\nonumber~,
\end{eqnarray}
where the functions $f({\bf k},{\bf p})$ and $g({\bf k},{\bf p})$ obey
the symmetry relation $f[g](-{\bf k},-{\bf q})=-f[g]({\bf k},{\bf
q})$.

To keep $\gamma_3$ to leading order in $1/S$ we retain only the
zeroth order propagators $G^0$ for each intermediate line. In
principle, $H_3$ allows for an additional two-particle irreducible
graph, with the incoming(outgoing) legs placed into the
particle-particle(hole-hole) channel and one intermediate line at
zero momentum and frequency. However, we verified that these
contributions vanish exactly.

Due to $\gamma_3$, Eqn. (\ref{bseqn}) is an integral equation with
respect to {\em both}, momentum and frequency. This is the first
major difference to Raman scattering from collinear HAFs, where only
$\gamma_4$ exists at $O(1/S)$. To proceed, further approximations
have to be made. Here we simplify $\gamma_3$ by assuming the
dominant contribution from the frequency summations to result from
the {\em mass-shell} of the propagators in the intermediate
particle-particle reducible sections of Fig. \ref{gamam}c)
\begin{eqnarray}
-i\omega_n & \approx & E_{\bf k} \nonumber \\
-i\omega_o & \approx & E_{\bf p}~.
\end{eqnarray}
This approximation will be best for sharp magnon lines and the
transferred frequencies $i\omega_m$ close to the van-Hove
singularities of $2 E_{\bf k}$.

In this approximation for  $\gamma_3$, the two-particle irreducible
vertex $\gamma$ simplifies to
\begin{eqnarray}\label{gamashell}
\gamma ({\bf k},{\bf p},\omega_n,\omega_o)\approx \gamma ({\bf
k},{\bf p}) =\gamma_3({\bf k},{\bf p})+\gamma_4 ({\bf k},{\bf p})
&&\\[.2cm]
=-\frac{1}{2S} \left( \frac{2 E_{{\bf k}-{\bf p}}f({\bf k},{\bf
p})g({\bf k},{\bf p})} {(E_{\bf k}-E_{\bf p})^2-E^2_{{\bf k}-{\bf
p}}} +h({\bf k},{\bf p})\right)~. &&\nonumber
\end{eqnarray}
Now we can perform the frequency summation over $\omega_o$ on the
right hand side of Eqn. (\ref{bseqn}) as well as the analytic
continuation $i\omega_m\rightarrow\Omega+i\eta\equiv z$. With this
$\Gamma$ in the latter equation turns into a function of ${\bf p}$
and $z$ only, leading to
\begin{eqnarray}
&&\sum_{\bf p} L_{{\bf k},{\bf p}}(z) \Gamma_{\bf
p}(z) = r^-({\bf k})
\label{lineqn1} \\
&&L_{{\bf k},{\bf p}}(z) = \delta_{{\bf k},{\bf p}} -
\frac{\gamma ({\bf k},{\bf p})}{z-2E^{(r)}_{\bf p}}~,
\label{lineqn2}
\end{eqnarray}
which is an integral equation with respect to momentum only. In the
rest of the paper the superscript '$r$' refers to the case
 when
{\em renormalized} propagators with $E^{r}_{\bf p}$ are taken in
the two-particle reducible part of the Raman intensity, while
$E_{\bf p}$  corresponds to the usage of {\em bare} propagators.

Close inspection of the vertex $\gamma ({\bf k},{\bf p})$ shows,
that it does not separate into a {\em finite} sum of products of
lattice harmonics of the triangular lattice. Therefore, Eqns.
(\ref{lineqn1})-(\ref{lineqn2}) cannot be solved algebraically in
terms of a finite number of scattering channels, but require a
numerical solution. On finite lattices this can be done by treating
Eqn. (\ref{lineqn1}) as a linear equation for $\Gamma_{\bf p}(z)$ at
fixed $z$. This marks another significant  difference between Raman
scattering from collinear and non-collinear antiferromagnets.

\begin{figure}
\centerline{\epsfig{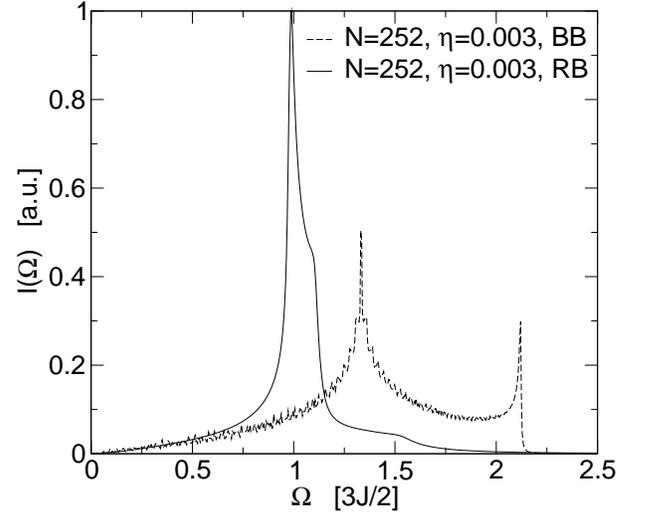}     } \caption{Raman
intensity neglecting final state interactions, i.e. replacing
$\Gamma$ by $M$ in eqn. (\ref{ramint2}). Scattering geometry:
$\phi=\theta=0$. Number of {\bf k}-points: $N\times N$. Dashed (BB):
bubble using bare magnon energies $E_{\bf k}$ from eqn. (\ref{ek}),
shown in top panel of fig. \ref{fig2}. Imaginary broadening
$\eta=0.003$ choosen such as to retain visible but small finite-size
oscillations. Solid (RB): bubble using renormalized magnon energies
$E^r_{\bf k}$ to $O(1/S)$ obtained from eqns. (10) - (12) of ref.
\cite{chubukov06} and shown in the middle and lower panel of fig.
\ref{fig2}. Finite-size oscillations are suppressed by $Im[E^r_{\bf
k}]$. The absolute scale of $I(\Omega)$ is set to unity, but the
relative scale of BB and RB is kept.} \label{raman1}
\end{figure}

Finally, the expression for the Raman intensity from Eqn.
(\ref{Fermi}) can be written as
\begin{eqnarray}\label{ramint}
I(\Omega) &=& const \times (J(\Omega)-J(-\Omega)) \\
\label{ramint2} J(\Omega) &=& Im \, [ \sum_{\bf k} \frac{M_{\bf
k}\,\Gamma _{\bf k}(\Omega+i\eta) }{\Omega+i\eta-2E^{(r)}_{\bf k}}
]~.
\end{eqnarray}

We now discuss the Raman intensity  for several levels of
approximations. First, we neglect final state interactions and set
$\Gamma_{\bf k}(z)\rightarrow M_{\bf k}$. Fig. \ref{raman1} shows
the Raman intensity as a function of the transferred photon
frequencies $\Omega$ for this case. This figure contrasts the Raman
bubble with bare propagators against that with renormalized ones.
Such results can be obtained on fairly large lattices, since they do
not involve a solution of the integral equation (\ref{lineqn1}), but
only a calculation of the one magnon self-energy \cite{chubukov06}.
We keep the shift $i\eta$ off the real axis  deliberately small in
this figure, in order to discriminate its effect from that of the
actual life-time broadening due to the imaginary part of $E^r_{\bf
k}$.

First we would like to note that we find the line shape to be
{\em insensitive} to the scattering geometry. This is in a sharp
contrast to Raman scattering from the square lattice HAF, where
Raman amplitudes in $A_{1g}$, $B_{1g}$ and $B_{2g}$ symmetries are
very different.

In case of the bare Raman bubble, one can see two  well-defined peaks, one
at energy $\Omega = 3/\sqrt{2}$ and one at $\Omega=4/3$ - both in units of
$3J/2$. These energies correspond to $2$ times that of the maximum and of
the BZ-boundary saddle-point of the classical spin-wave spectrum $E_{\bf
k}$. Clearly, the dominant spectral weight does not stem from the {\bf
k}-points at the upper cut-off of the linear spin-wave energy but from the
BZ-boundary. This does not reflect the bare two-magnon density of states
but is an effect of the Raman matrix element $M_{\bf k}$, which samples the
BZ regions preferentially.

Switching on $1/S$ corrections, two modifications of
the intensity occur. First, both maxima are shifted downwards by a
factor of $\sim 0.7$ due to the corresponding renormalizations of
the one magnon energies. Second, as the BZ-boundary saddle-point of
$E_{\bf k}$  has turned into a flat region, occupying substantial parts
of the BZ for $E^r_{\bf k}$, the intensity of the lower energy peak
is strongly enhanced due to the very large density of one-magnon
states. Equally important, the imaginary part of $E^r_{\bf k}$ is
finite in the BZ-region which corresponds to the maximal one-magnon
energies. This smears the peak at the upper frequency cut-off in
$I(\Omega)$ almost completely - as can seen from the solid line in
Fig. \ref{raman1}. In contrast to that, $Im[E^r_{\bf k}]$ almost
vanishes in the flat regions on the BZ-boundary due to phase-space
constraints \cite{mike}, leading to a further relative enhancement
of the intensity from there.

\begin{figure}
\centerline{\epsfig{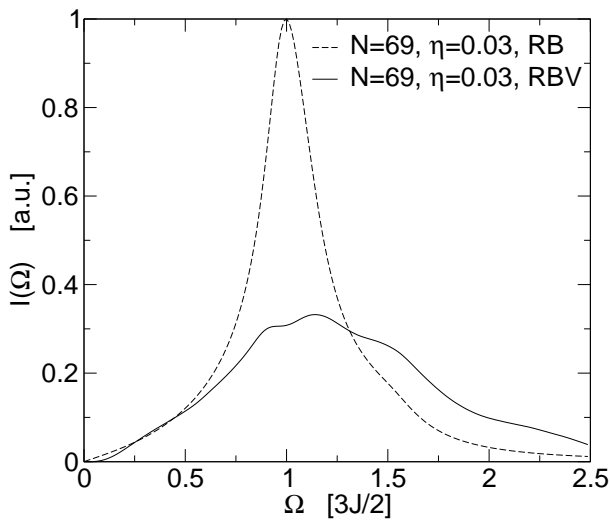}} \caption{Effect of
final state interactions on Raman intensity. Scattering geometry:
$\phi=\theta=0$. Number of {\bf k}-points: $N\times N$. Imaginary
shift off real axis is $\eta = 0.03$. Dashed line (RB): replacing
$\Gamma$ by $M$ in Eqn. (\ref{ramint2}) and using bubble with
renormalized magnon energies $E^r_{\bf k}$ to $O(1/S)$ obtained from
Eqns. (10) - (12) of Ref. \cite{chubukov06}. Solid line (RBV): using
dressed vertex $\Gamma$ obtained from Eqn. (\ref{lineqn1}) in Eqn.
(\ref{ramint2}) and renormalized magnon energies $E^r_{\bf k}$. The
absolute scale of $I(\Omega)$ is set to unity, but the relative
scale of RB and RBV is kept.} \label{raman2}
\end{figure}

Next we turn to final-state interactions. In Fig.\ref{raman2}, we
compare $I(\Omega)$ from the Raman bubble obtained with propagators
renormalized to $O(1/S)$ and only bare Raman vertices to the
intensity obtained by including also the dressed Raman vertex
$\Gamma_{\bf k}(z)$ from Eqn. (\ref{lineqn1}). The numerical
solution of the latter equation requires some comments. Since the
kernel $L_{{\bf k},{\bf p}}(z)$ is not sparse and has rank $N^2
\times N^2$, already moderate lattice sizes lead to rather large
dimensions and storage requirements for the linear solver. We have
chosen $N=69$, leading to a 4761$\times$4761 system which we have
solved 200 times to account for 200 frequencies in the interval
$\Omega\in [0,2.5]$. The kernel has points and lines of singular
behavior in $({\bf k},{\bf p})$-space, which stem from the
singularities of the Bogoliubov factors $u[v]({\bf k})$ in
$f[g][h]({\bf k},{\bf p})$ and from the energy denominators in Eqn.
(\ref{gamashell}). In principle, such regions are of measure zero
with respect to the complete $({\bf k},{\bf p})$-space, yet we have
no clear understanding of their impact on a solution of Eqn.
(\ref{lineqn1}) as $N\rightarrow\infty$. In our case, i.e. a finite
lattice, we have chosen to regularize these points and lines by
cutting off eventual singularities in $L_{{\bf k},{\bf p}}$. The
comparatively small system sizes require a rather larger artificial
broadering $i\eta$ in order to achieve acceptably smooth line
shapes. This can be seen by contrasting the dashed curve in Fig.
\ref{raman2}  and solid curve in Fig.\ref{raman1}, which correspond
to {\em identical} quantities, however   for different finite
systems, $252\times 252$ vs $69\times 69$.

The main message put forward by Fig. \ref{raman2} is that the final
state interactions lead to a flattening of the peak from the
two-magnon density of states, transforming it to a rather broad
Raman continuum. This can be understood, at least qualitatively,
from the RPA-like functional form of the Bethe-Salpeter equation.
Discarding momentum dependencies and iterating the two-magnon bubble
times the irreducible vertex $\gamma$, leads to a renormalization of
the intensity by a factor, roughly of the form $\sim
1/(1-\gamma\cdot\rho(\Omega))$, where $\rho(\Omega)$ refers to the
two-magnon bubble. Directly at the peak position of the Raman bubble
this renormalization factor may get small, thereby suppressing the
over-all intensity. While exactly the same mechanism is at work also
for the square lattice HAF, its impact on the spectrum is complete
different. In the latter case the peak-intensity without final state
interactions is at the upper cut-off of the Raman intensity.
Suppression of this peak-intensity simply shifts the maximum
intensity to lower frequencies within the Raman spectrum. This shift
is then interpreted in terms of a two-magnon binding energy. Such
reasoning cannot be pursued in the present case.

\section{Conclusion and Discussion}\label{sec5}

To summarize, we have investigated magnetic Raman scattering from the
two-dimensional triangular Heisenberg antiferromagnet considering various
levels of approximation within a controlled $1/S$-expansion. Our study has
revealed several key differences as compared to the well-known magnetic
Raman scattering from the planar square lattice spin-$1/2$ antiferromagnet.

First, we found that the intensity profile is insensitive to the
in-plane scattering geometry of the incoming and outgoing light at
$O(1/S)$. This has to be contrasted against the clear difference between
A$_{1g}$ and B$_{1g,2g}$ symmetry for the square-lattice case.

Second, on the level of linear spin-wave theory, we showed that the
Raman intensity has two van-Hove singularities. The less intensive
peak is located at the upper edge of the two-magnon density of
states and stems from twice the maximum of the one-magnon energy.
This is similar to the square lattice case. However, the dominant
peak is located approximately in the center of the two-magnon
density of states. This peak stems from the Loudon-Fleury
Raman-vertex strongly selecting the Brillouin zone boundary regions
where the one-magnon dispersion on the triangular lattice has an
additional weak van-Hove singularity. This is absent on the square
lattice.

Next, we calculated the Raman intensity with the one-magnon
spectrum, renormalized to $O(1/S)$, however neglecting final-state
interactions within the Raman process. In this case we have obtained
a sharp and almost $\delta$-functional Raman peak at energy $\sim 3
J/2$. At this energy the real part of the renormalized one-magnon
dispersion shows a large plateau-region at the Brillouin zone
boundary with a roton-like shallow minimum. Moreover, due to
phase-space constraints the one-magnon life-time is large in this
region. Therefore, the two-magnon density of states in this region
is strongly enhanced, as compared to the linear spin-wave result. In
contrast to that, the intensity at the upper edge of the spectrum is
suppressed further, since the $O(1/S)$ corrections lead to the
significant one-magnon damping. Finally the overall width of the
spectrum is reduced by a factor of approximately $0.7$.

In a last step, we  considered the impact of the final-state
interactions to $O(1/S)$. Due to the non-collinear ordering on the
triangular lattice, and in sharp contrast to the square-lattice
case, we find, that even to lowest order the two-magnon scattering
is neither instantaneous in time, nor separable in momentum space.
Our solution of the corresponding Bethe-Salpeter equation reveals a
broad continuum-like Raman profile which results from a smearing of
the intensity of the two-roton peak by virtue of repeated two-magnon
scattering. While, at this order in $1/S$ the over-all form of the
Raman profile is reminiscent of that on the square-lattice, one has
to keep in mind, that in the latter case the position of the maximum
in the center of the Raman continuum has to interpreted rather
differently, namely in terms of a two-magnon binding effect.

In conclusion we hope that our theoretical investigation will stimulate
further experimental analysis of triangular, and more generally frustrated
magnetic systems by Raman scattering. Several novel materials with
triangular structure have been investigated thoroughly over the last few
years, among them the cobaltites, Na$_x$CoO$_2$ \cite{cob}, and the
spatially anisotropic triangular antiferromagents Cs$_2$CuCl$_4$ \cite{Cs}
and $\kappa$-(BEDT-TTF)$_2$Cu$_2$(CN)$_3$ \cite{Shimizu2003}.
To our knowledge however, magnetic Raman scattering on such systems remains
a rather open issue.

\section{Acknowledgements}
We would like to thank A. Chubukov for useful discussions. One of us
(W.B.) acknowledges partial support by the DFG through Grant No. BR
1084/4-1 and the hospitality of the KITP, where this research was supported
in part by the NSF under Grant No. PHY05-51164.

\end{document}